# Efficient hot electron generation via low-coherence lasers

Huiya Liu,[1,*] Yao Zhao,[2,*] Ning Kang,[1,†] Fujian Li,[3] Guoxiao Xu,[3] Honghai An,[3] Jun Xiong,[3] Zhiyong Xie,[3] Xichen Zhou,[3] Zhiheng Fang,[3] Wei Wang,[3] Lailin Ji,[3] Xiaohui Zhao,[3] Liang Hao,[4] Lifeng Wang,[4] and Anle Lei[3,‡]

[1] *Joint Laboratory of High Power Laser and Physics, Shanghai Institute of Optics and Fine Mechanics, Chinese Academy of Sciences, Shanghai 201800, China*

[2] *School of Science, Shenzhen Campus of Sun Yat-sen University, Shenzhen 518107, China*

[3] *Shanghai Institute of Laser Plasma, China Academy of Engineering Physics, Shanghai 201800, China*

[4] *Institute of Applied Physics and Computational Mathematics, Beijing 100094, China*

**Abstract:** Hot electrons generated in laser-produced plasmas are a central focus in inertial confinement fusion, laboratory astrophysics, and high-energy-density physics. These electrons originate from instabilities in nonlinear laser-plasma interactions, which are critically modulated by laser bandwidth. Here, we experimentally demonstrate enhanced generation of hot electrons by utilizing instantaneous low-coherence lasers with two bandwidths (0.2% and 0.6%) at intensities of $2–8\times10^{14}$ W/cm$^2$ and energies up to 620 J. A significant enhancement of hot electron temperature and hard X-ray yield is observed with the broadband lasers compared to a conventional narrowband laser. The results show that the hot electron energy conversion efficiency of the 0.6% broadband laser is approximately 4 times higher than that of the narrowband laser, reaching a maximum value of 2.8%. These findings validate a moderate-bandwidth laser as an efficient hot electron source and support the generation of bright X-ray sources for advanced imaging in high-energy-density physics.

## I. Introduction

The generation of hot (suprathermal, kinetic energy above 30 keV) electrons in laser-plasma interactions has long been crucial for the development of the laser-driven inertial confinement fusion (ICF), laboratory astrophysics, and high-energy-density physics (HEDP) [1–5]. In conventional central hot-spot ignition, hot electrons can deposit energy into cold fuel layers, inducing preheating that degrades fuel compression during implosion [3]. Conversely, in shock ignition [6,7], hot electrons can amplify the shock strength via amplified resistive heating, thereby improving energy coupling efficiency and ignition performance [8]. In extreme matter studies relevant to astrophysics and HEDP, hot electrons play critical roles in a range of applications, including heating dense plasmas to determine equations of state, generating intense magnetic field for studying high-energy particle acceleration in universe [9,10], and producing bright X-rays for radiography of high-energy-density plasmas [11].

Laser-produced hot electrons are accelerated by electron plasma waves (EPWs), which are generated by the parametric instabilities in the plasma corona [12,13]. Two particularly important instabilities are the stimulated Raman scattering (SRS) [13–16] and the two-plasmon decay (TPD) [17–19]. As the growth of such instabilities relies on the resonance between laser and plasma waves, adjusting the laser bandwidth, which affects the laser coherence time, may modify the resonant processes, thereby influencing the production of hot electrons [20–22]. Conventional lasers at large facilities used in ICF and HEDP research are usually narrowband, corresponding to a fractional bandwidth less than 0.01% [23]. Based on new advances in broadband laser techniques, several broadband laser systems with moderate (0.1%–1%) or large (> 1%) bandwidths have been developed worldwide, such as Kunwu laser at Shanghai Institute of Laser Plasma in China (up to 0.6%) [24,25], FLUX laser at University of Rochester in the US (up to 1.5%) [26], and PHELIX broadband laser at Helmholtzzentrum für Schwerionenforschung in Germany (up to 0.5%) [27]. Such developments pave the way for further research on controlling hot-electron production by modifying laser bandwidth. Recent experimental studies on the Kunwu laser [28–31], found that the stimulated Brillouin scattering (SBS) shows robust mitigation, while SRS becomes more intense under moderate bandwidths at relatively high intensities [32,33]. Similar results were also observed in recent PHELIX laser experiments, with enhanced hot-electron generation measured [34].

In this paper we report on the experimental results on hot electrons generated by lasers with two different moderate bandwidths (0.2% and 0.6%), which are compared to the conventional narrowband case to assess the impact of moderate bandwidth on hot-electron production. The electron spectrometer (ESM) and the hard X-ray spectrometer were used to obtain temperature and number of hot electrons. Results show that both the temperature and number of hot electrons are enhanced with the broadband lasers compared to those with a narrowband laser, and the enhancement of the 0.6% bandwidth is more significant than that of the 0.2% bandwidth. The energy conversion efficiency of hot electrons with respect to the incident laser is as high as 2.8% for the 0.6% bandwidth, which is about 4 times higher than that for the narrowband case. Therefore, low-coherence lasers can lead to significantly enhanced hot-electron production and show great potential for improving X-ray source brightness, which is essential for advanced imaging applications in laboratory astrophysics and HEDP research.

## II. Experimental setup

The experiment was conducted on the Kunwu laser, which uses a superluminescent diode as the seed source and produces an instantaneous broadband laser output with a maximum bandwidth of 0.6%. The experimental setup is schematically shown in Fig. 1(a). The laser beam was focused perpendicularly onto a 50 μm-thick $C_8H_8$ target (density: 1.05 g/cm$^3$) with a 15 μm-thick titanium (Ti) layer on the back. A continuous phase plate (CPP) was used to form a relatively uniform irradiance on the target surface with 180 μm in diameter. Up to 620 J of laser energy, at the second harmonic (0.53 μm) and in a 3.2 ns square pulse, was delivered in the experiments, resulting in a maximum intensity of approximately $8\times10^{14}$ W/cm$^2$. The spectra of the broadband and narrowband lasers are shown in Fig. 1(b), where the factional bandwidths ($\Delta\omega/\omega_0$) for the broadband lasers are approximately 0.6% and 0.2%, while that for the narrowband laser is less than 0.01%.

Electron energy spectra in the range of 20–400 keV were measured using ESMs from both the forward and backward directions with respect to the laser incidence. These ESMs employ a permanent magnetic field for dispersion [35], and such devices are widely used to measure escaped hot electron energy spectra and further estimate hot electron temperature [36-39]. A Ross pair type hard X-ray spectrometer (HXSM) [40,41] was placed in the laser forward direction to measure the forward emission of 5–88 keV bremsstrahlung X-rays associated with the hot electrons. A 0.45 T magnet was located in front of the HXSM to prevent electrons with energies up to 1 MeV from entering the spectrometer. Two fiber probes coupled with 2 spectrometers were used to measure the scattered light spectra from the backward direction with a diffuser (SBS) and side direction (SRS), respectively. The laser spot on the target was recorded by an X-ray pinhole camera (XPHC). All these diagnostics are located in the equatorial plane of the experiment, and the laser polarization direction is perpendicular to this plane.

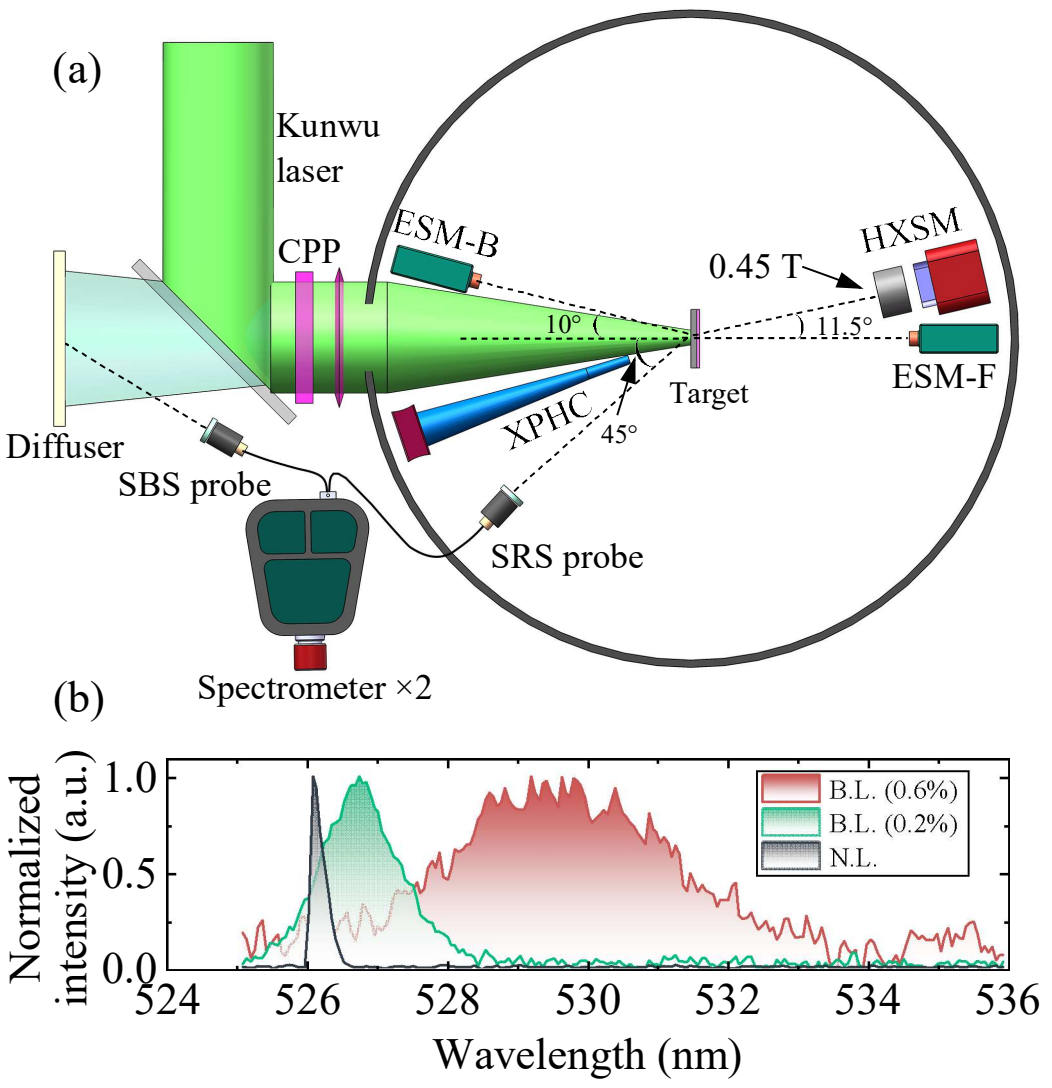

**Figure 1**. Experimental setup and laser spectra. (a) Schematic diagram of the experimental setup. The ESM denoted with the suffixes -F and -B refer to the electron spectrometer measuring the forward and backward electron spectrum, respectively. (b) Spectra of the broadband (B.L.) lasers and the narrowband (N.L.) laser.

## III. Results

### A. Hot electron energy spectra

Experimental results show clear influence of laser bandwidth on hot-electron generation. Fig. 2(a) presents typical electron kinetic energy spectra measured with the ESMs for the broadband and narrowband lasers with energy ~500 J. Spectra were fitted with a 3-D Maxwellian distribution $f(E_e) = 2/\sqrt{\pi} * N_0 T_{hot,e}^{-3/2}\sqrt{E_e}\exp(-E_e/T_{hot,e})$ [42], where $N_0$ is the electron number, $T_{hot,e}$ the electron temperature, and $E_e$ the kinetic energy of an electron. The fitted $T_{hot,e}$ for each shot are shown in Fig. 2(a). $T_{hot,e}$ for the backward electron spectra, as measured by ESM-B, are 39.56 keV for the narrowband laser, 56.84 keV for the 0.2% broadband laser, and 63.36 keV for the 0.6% broadband laser. Meanwhile, $T_{hot,e}$ for the forward electron spectra measured by ESM-F are 49.53 keV, 59.23 keV, and 75.59 keV, respectively. The electron temperature of the 0.6% and 0.2% broadband lasers is higher than that of the narrowband laser, for both the forward and backward measurements. Besides, for each shot, the forward electron spectrum exhibits higher temperature compared to the backward spectrum for both the broadband and narrowband lasers. This can be attributed to spectral hardening during electron transport through the target, where lower-energy electrons are preferentially lost, reshaping the transmitted spectrum toward higher energies. This finding is confirmed by subsequent Monte Carlo simulations (see Fig. 4(a)), where the temperature of electrons emitted from the target rear also increases accordingly. Fig. 2(b) shows $T_{\mathrm{hot,e}}$ for the backward electron spectra versus laser intensity, in which $T_{\mathrm{hot,e}}$ is about 60–80 keV for the 0.6% broadband laser at intensities of 2–8×10$^{14}$ W/cm$^2$, roughly 2 times that for the narrowband laser.

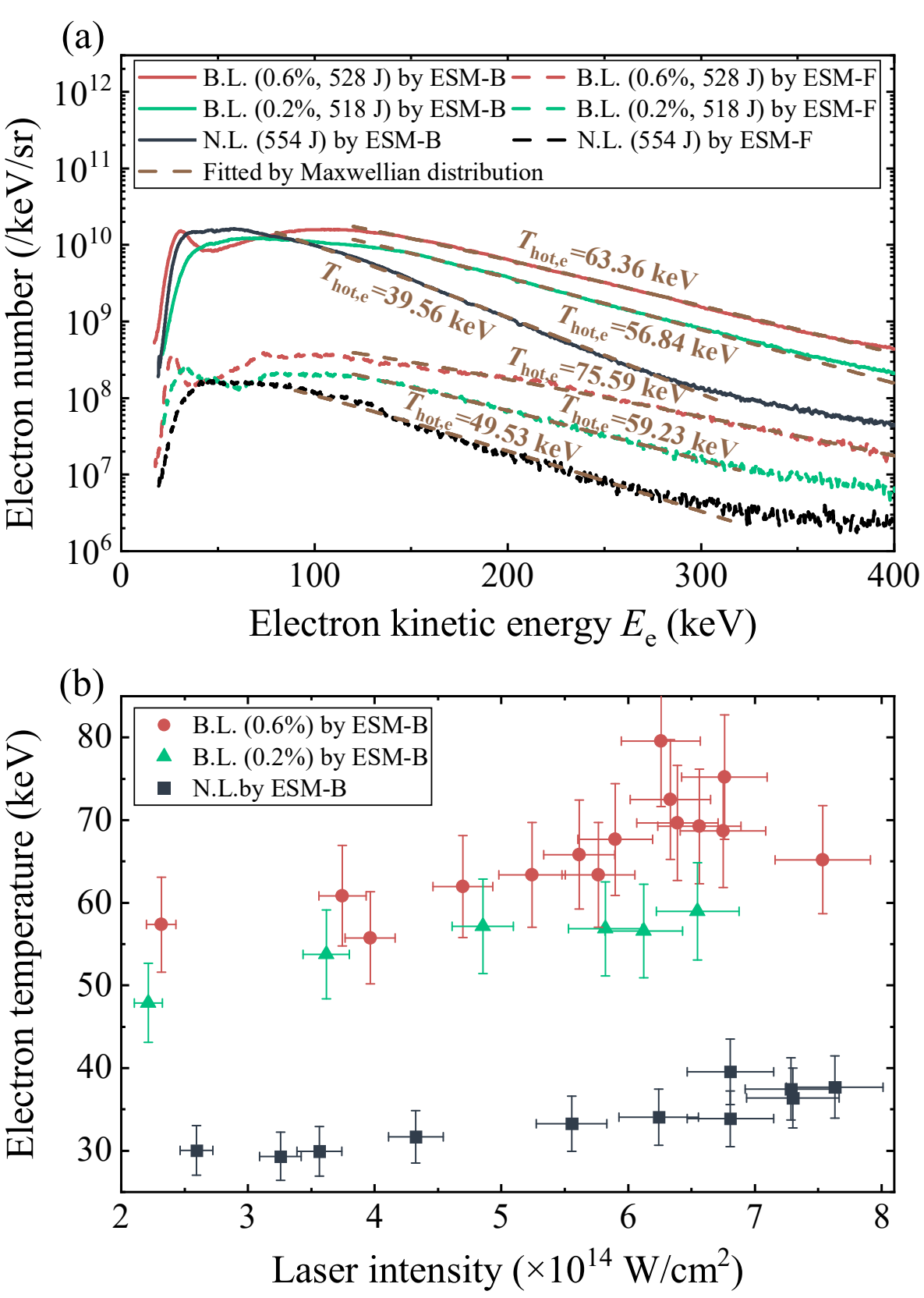

**Figure 2**. Electron energy spectra and electron temperature. (a) Typical electron energy spectra measured with ESM-F and EMS-B. The dashed lines are the fitted curves by a 3-D Maxwellian distribution. (b) Plot of electron temperature versus laser intensity for the broadband and narrowband lasers.

## B. Hard X-ray spectra

Fig. 3(a) presents the X-ray spectra measured with the HXSM from the same shots as shown in Fig. 2(a). In these X-ray spectra, radiation tails from the thermal plasma are contained and dominated in the lower energy range. The X-rays with energies over 20 keV are mainly produced by hot electrons through bremsstrahlung. The fraction of laser energy ($E_{\mathrm{laser}}$) converted to the X-ray energy in the range of 19.1–88 keV ($E_{\mathrm{h,X-ray}}$) is calculated, as shown in Fig. 3(b). The enhancement of hard X-ray generation by using broadband lasers is significant. The 0.6% broadband laser yields approximately 5 times the hard X-rays of the narrowband laser, while the 0.2% broadband laser yields approximately 2 times as much.

## C. Hot electron production

The number of hot electrons measured in vacuum is heavily reduced due to the Alfven limit of current [43], meaning that most hot electrons generated inside the target cannot escape into vacuum. The number of hot electrons generated can be inferred from their bremsstrahlung emission. To estimate the total number of hot electrons in the experiment, we perform Monte Carlo (MC) simulations with the *EGS*nrc code [44], using input parameters from ESM and HXSM measurements. The forward flying electrons impinging on the plasma may either be scattered backward or decelerated causing bremsstrahlung radiation [42], which correspond to the measured backward electron energy spectrum and hard X-ray spectrum, respectively. Therefore, we used the backward electron spectrum measured by ESM-B as an input to go through the target in the MC simulation. As discussed before, the scattered electrons collected by ESM-B are escaping hot electron, which account for only a small fraction of the generated hot electrons, so we followed refs. [39,45,46] and used the measured electron energy spectra not in quantity but in spectral distribution as an input. Then, the output hard X-ray spectrum was compared with the experimental result to validate the simulation.

An example of the simulation results is presented in Fig. 4(a), wherein the photon number of the simulated Bremsstrahlung spectrum is normalized to the intensity of the fourth channel of the HXSM with an energy range of 29.3–50.2 keV. This particular channel was selected to compromise between the influence of plasma thermal radiation on lower pairs and the insufficient intensity on higher pairs. Through this normalization process, the total electron number ($N_{\mathrm{e}}$) required to generate the specified intensity of Bremsstrahlung radiation can be determined. In Fig. 4(a), the output electron spectrum at the target backside obtained from the MC simulation is scaled by a factor of $N_{\mathrm{e}}/N_{\mathrm{MC}}$ to facilitate comparison with the experimental electron spectrum measured by ESM-F, where $N_{\mathrm{MC}}$ refers to the number of electrons input to the Monte Carlo (MC) simulation. One finds that the fitted temperature of the output forward flying hot electrons in the MC simulation (72.97 keV) is roughly the same as that measured by ESM-F in the experiment (75.59 keV), which also validate the simulation in addition to the comparison in the hard X-ray spectra. However, the output forward flying hot electrons in the simulation has a significant larger number than that measured by ESM-F in the experiment, which may be attributed to the influence of the magnetic field on the transport of hot electrons in the experiment that is not included in the MC simulation. For instance, hot electrons may have longer travel distances in a magnetic field before they penetrate the target area, resulting in higher losses in the experiment than that in the simulation. Consequently, the total energy of hot electrons ($E_{\mathrm{hot,e}}$) can be calculated by $E_{\mathrm{hot,e}} = N_{\mathrm{e}}\varepsilon_0\, \eta_{\mathrm{sim}}/\eta_{\mathrm{exp}}$, where $\varepsilon_0$ denotes the normalized input spectrum energy for a single electron, $\eta_{\mathrm{sim}}$ and $\eta_{\mathrm{exp}}$ are the hot electron loss rate in the simulation and experiment, respectively.

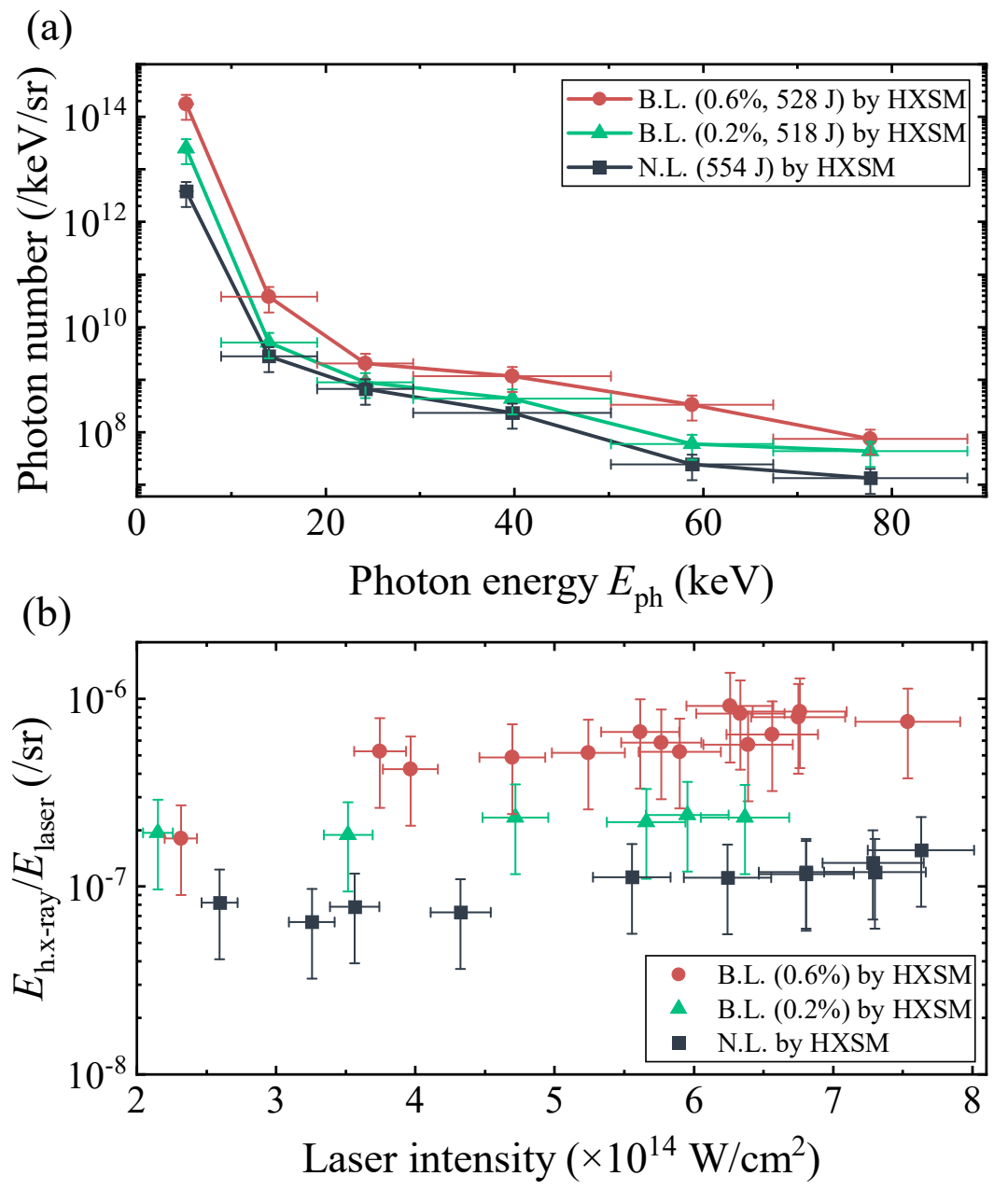

**Figure 3**. X-ray spectra and conversion efficiency. (a) X-ray spectra from the same shots as shown in Fig. 2(a) measured by HXSM. Error bars are caused by the width of each spectrometer channel and intensity calculation. (b) Fraction of laser energy ($E_{\mathrm{laser}}$) converted into the hard X-ray bremsstrahlung emission energy ($E_{\mathrm{h,X-ray}}$, 19.1–88 keV) as a function of incident laser intensity.

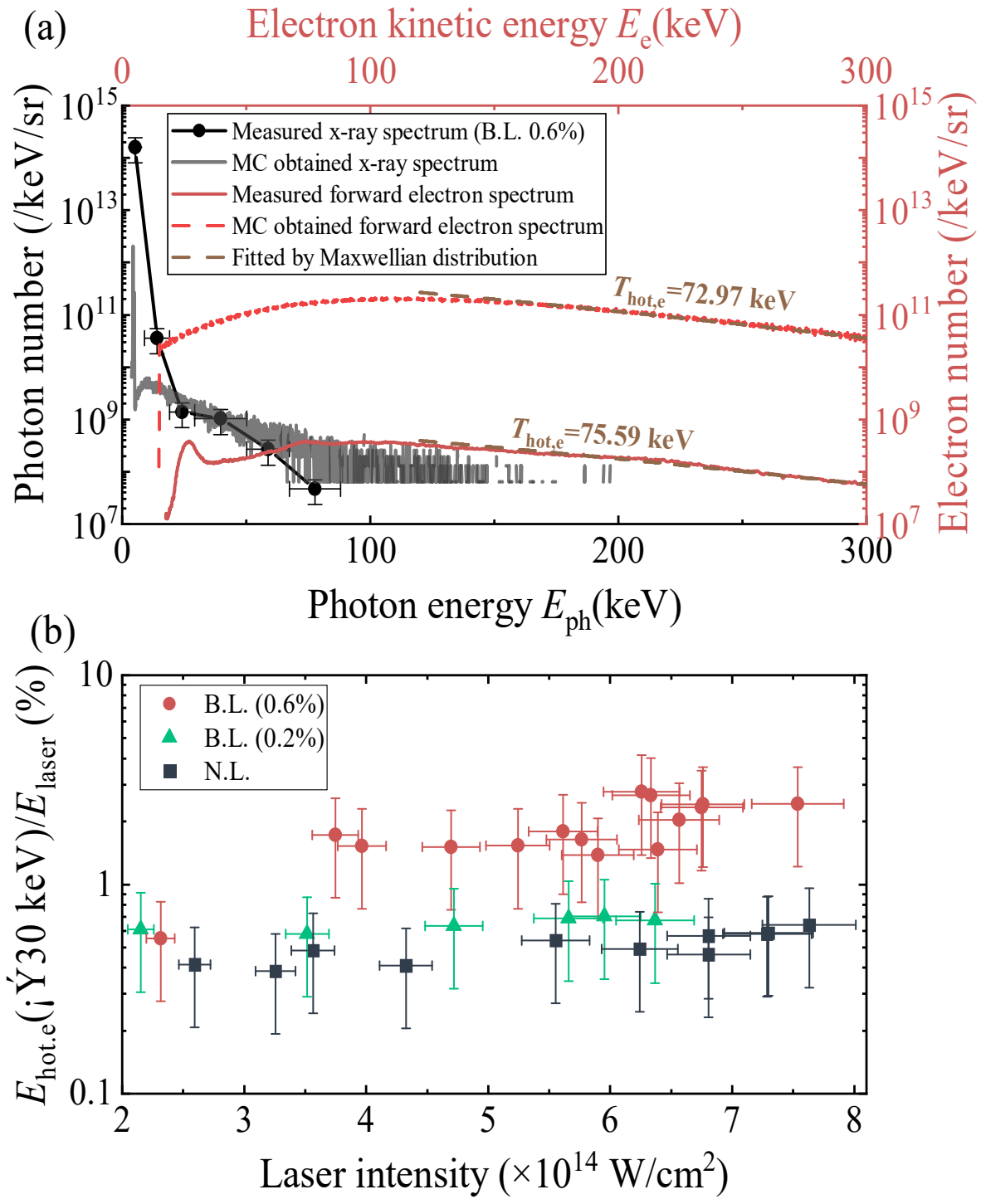

**Figure 4**. MC simulation results and hot electron energy conversion efficiency. (a) Comparison between the MC simulation results and the experimental results for the 0.6% broadband shot at 528 J in Fig. 2(a). (b) The fraction of laser energy ($E_{\mathrm{laser}}$) converted to the total energy of hot electrons ($E_{\mathrm{hot,e}}$) with kinetic energy exceeding 30 keV. The uncertainty originates from the uncertainty in the hard X-ray data.

Using the above method, the fraction of laser energy converted to the total hot electron energy ($\geqslant$30 keV) is given, as shown in Fig. 4(b). The conversion efficiencies at a laser intensity of $(6\pm0.5)\times10^{14}$ W/cm$^2$ are 0.52% for the narrowband laser, 0.69% for the 0.2% broadband laser, and 2.07% for the 0.6% broadband laser, respectively. The 0.6% bandwidth case shows approximately 4 times as high as the narrowband case, whose conversion efficiency peaks at 2.8%. Meanwhile, the energy fraction of hot electrons for the 0.2% bandwidth case exhibits a less pronounced increase compared to the narrowband laser. As Bremsstrahlung radiation intensity follows $I_{\mathrm{brem}} \propto E_{\mathrm{hot,e}} \times \exp\left(-h\nu/T_{\mathrm{hot,e}}\right)$ [12], higher hot electron temperatures at fixed kinetic energy lead to enhanced hard X-ray production. This mechanism explains why the hard X-ray conversion efficiency of the 0.2% broadband laser in Fig. 3(b) is nearly 2 times that of the narrowband laser.

## IV. Discussion

The experimental results demonstrate a clear enhancement in hot electron temperature and number with the broadband lasers, particularly for the 0.6% bandwidth case. Since these electrons are primarily attributed to parametric instabilities in the plasma corona, the light spectra of SBS and SRS were simultaneously measured in the experiment, as shown in Fig. 5. The spectra in Fig. 5(a) were identified as SBS backscatter, which include contributions from laser reflection, as they can hardly be experimentally distinguished. Notably, the 0.6% bandwidth case shows an approximate 50% reduction compared to the narrowband laser. The reduction of SBS considerably improves laser absorption, benefiting the efficient production of hot electrons.

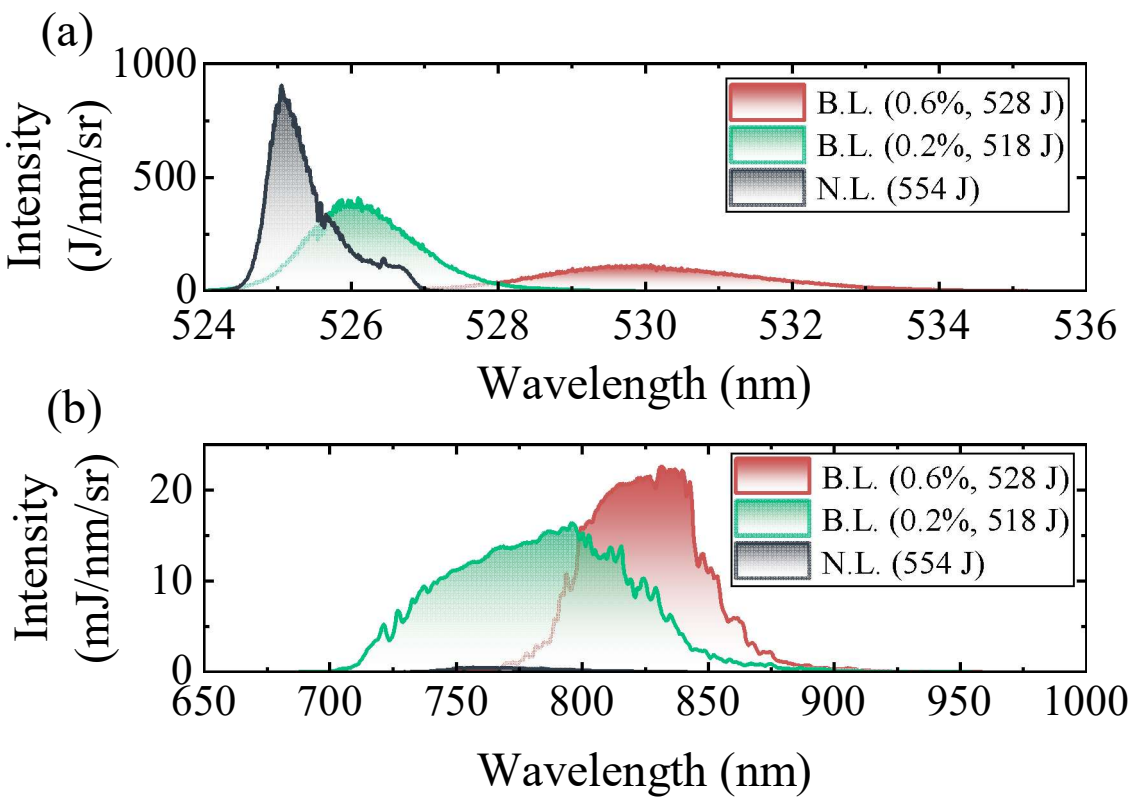

**Figure 5**. Measured scattering spectra**.** (a) The SBS backscatter spectra including contributions from laser reflection. (b) The SRS side scatter spectra measured at 45° relative to the laser beam.

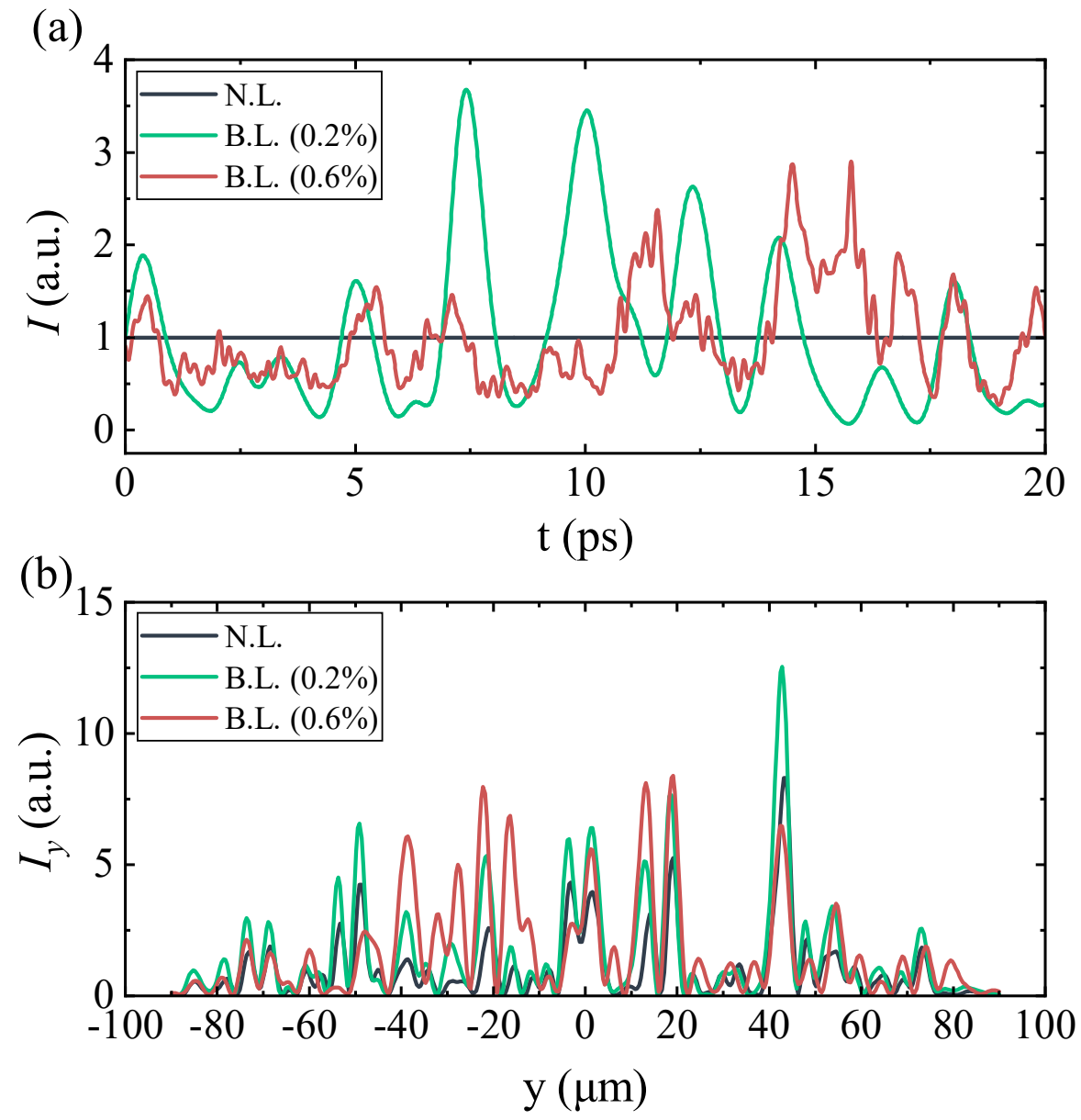

**Figure 6**. Spatiotemporal distributions of laser intensity used in the 2-D PIC simulations. (a) Laser temporal intensity distribution. (b) Laser spatial intensity distributions at $t = 18$ ps. The temporal modulation of the broadband laser enables localized intensity enhancement by up to tenfold relative to the average intensity at some spatial positions.

Fig. 5(b) presents scatter spectra measured at 45° relative to the laser incidence, showing significantly enhanced side scatter of SRS for the broadband lasers. The scattering light with longer wavelengths was observed for the laser with 0.6% bandwidth compared to the 0.2% case, which implies SRS can be developed at higher plasma densities for the laser with a larger bandwidth. The phase velocity of EPW increases with plasma density, which heats electrons to a higher kinetic energy. The enhanced SRS scatter may attribute to both the mitigated competition from SBS and the intrinsic bandwidth effects on SRS. Critically, the latter one may play a dominant role in the generation of hot electrons. To investigate these underlying mechanisms, we conducted a series of numerical simulations using experimental parameters.

The case of an incident energy of 500 J is discussed as an example for the following analyses. Radiation hydrodynamic simulations have been performed to estimate the plasma conditions by use of the FLASH code [47,48], where the electron and ion temperatures are ~1.1 keV and ~0.5 keV in underdense plasma regions. The 2-D PIC simulations were conducted with the EPOCH code [49]. An exponential plasma density profile $n_e(x)/n_{\mathrm{c}} = 0.067\exp[(x/\lambda_0 - 10)/375]$ is adopted for the simulations, with a relatively broad density range of $[0.067,\ 0.31]\ n_{\mathrm{c}}$, where $n_{\mathrm{c}}$ is the critical density. To cover this density range, a large simulation domain of 300 μm × 54 μm is adopted.

All the incident lasers are smoothed by a CPP, which has been considered in the presented simulations [50]. Figs. 6(a) and 6(b) display the temporal and spatial intensity distributions of the laser used in the simulations, which contains the major characteristics of the broadband lasers. Due to the temporal intensity fluctuation of the broadband laser, the intensity in some speckles induces an order-of-magnitude intensity enhancement.

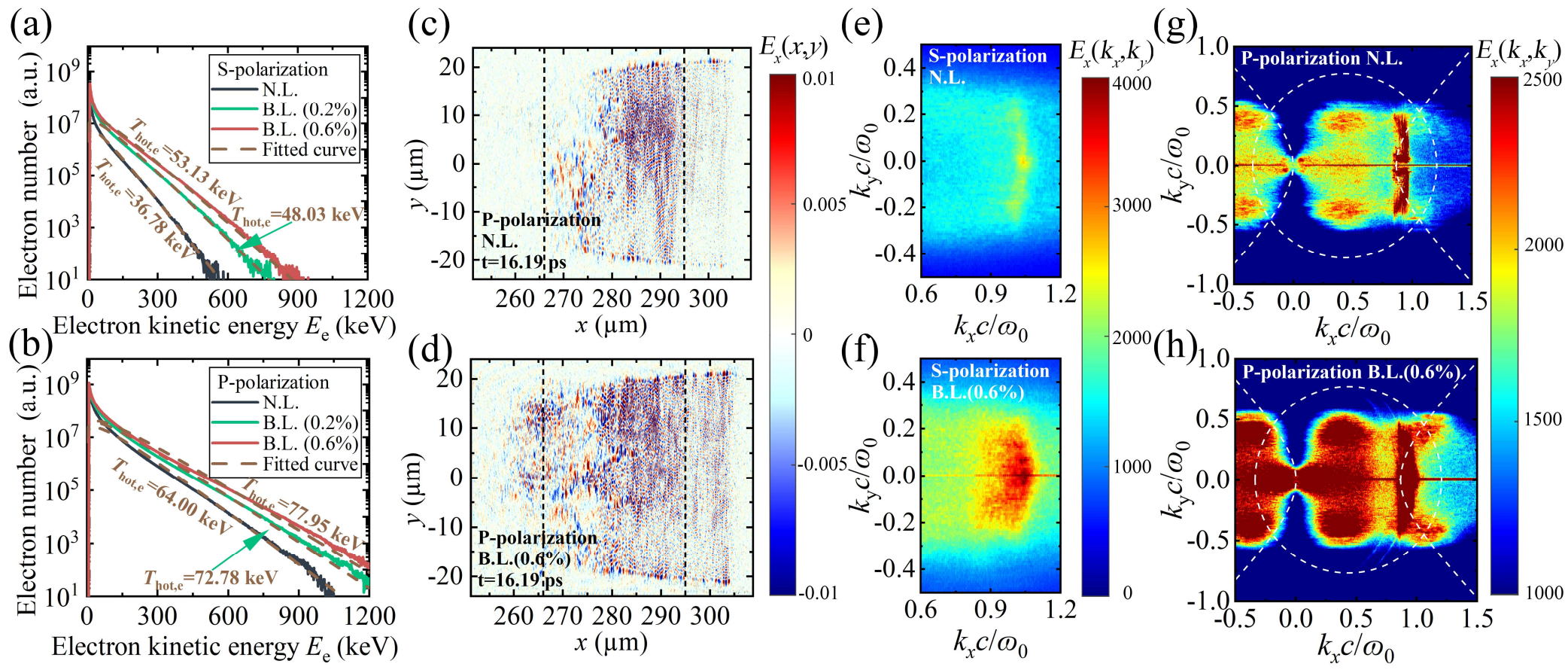

**Figure 7**. Two-dimensional PIC simulation results. Electron energy spectra in the time window [0.70, 17.57] ps driven by s-polarized (a) and p-polarized (b) laser beams, respectively. (c, d, g, h) Results for p-polarized cases. (e, f) Results for s-polarized cases. (c) and (d) show spatial distributions of the electrostatic field driven by a narrowband and a broadband laser at t = 16.19 ps, respectively. Vertical dashed lines in (c, d) indicate the initial density ranges [0.25, 0.29] $n_c$. (e, f) Wavevector distributions of the electrostatic field excited by s-polarized lasers, integrated over the time window [8.78, 17.57] ps for the narrowband and broadband laser, respectively. (g, h) Corresponding distributions for p-polarized lasers, integrated over [8.78, 17.57] ps. The white dashed hyperbola and circle denote the expected wavenumber distribution of TPD and the dispersion relation of electrostatic waves, respectively.

In PIC simulations, 2D models with separate s-polarization (electric field perpendicular to the simulation plane) and p-polarization (parallel to the plane) were used to approximate 3D normal laser-plasma interaction results. S-polarized lasers induce stronger overall SRS than p-polarized ones, as their incident and scattered waves share the same polarization angle [51]. Conversely, p-polarized lasers possess an in-plane electrostatic component that enables TPD excitation. The average laser intensity employed in our simulations is about $6\times10^{14}$ W/cm$^2$ for the 527 nm incident laser, which is well above the absolute thresholds for both SRS ($1.1\times10^{14}$ W/cm$^2$) and TPD ($7.4\times10^{13}$ W/cm$^2$) [21], ensuring that both instabilities are strongly excited.

Electron energy spectra shown in Figs. 7(a) and 7(b) demonstrate that the number and temperature of hot electrons generated by p-polarized lasers are much higher than that of s-polarized lasers, which leads to smaller discrepancies of hot electrons between broadband and narrowband lasers for the p-polarized simulation results. The experimental results are intermediate between the p-polarized and s-polarized cases, with the broadband data being marginally closer to the p-polarized simulations. In comparison, narrowband data correspond closely to s-polarized simulations in terms of electron temperature.

The simulation results shown in Figs. 7(c) and 7(d) indicate that a more intense electrostatic field is excited by the broadband lasers. A moderate bandwidth of 0.6% favors the growth of absolute instabilities around $0.25\,n_c$, where the multi-color components are strongly coupled [52,53]. Correspondingly, hot electrons are predominantly generated in this region, with the electrostatic field exhibiting substantially higher intensity and phase velocity than convective modes. Meanwhile, the transient intensity of broadband lasers caused by the envelope fluctuations is almost 2 times as high as the average intensity of narrowband laser, which leads to the more intense absolute instability modes [30,53]. For a narrowband beam, there exists a finite density range for the development of absolute instabilities around $0.25n_c$ [52]. As an intrinsic effect of a broadband laser, the density region of absolute modes is broadened by the moderate bandwidth, which can be observed from the Fig. 7(d). Therefore, much more electrons are heated by the 0.6% broadband laser in a larger scale of space.

As another intrinsic effect of a broadband laser, the wavevector width of electrostatic field excited by absolute SRS is determined by the bandwidth of incident laser, that is $\delta k_x \sim \delta k_0$, where $k_x$ and $k_0$ are the wavevectors of electrostatic field and incident laser, respectively. The wavevector distributions of the electrostatic wave, obtained from s-polarized simulations and shown in Figs. 7(e) and 7(f), indicate that more intense absolute instability with a wider wavevector spectrum is developed by the broadband laser. Electrons trapped by the intense electrostatic field can be accelerated continuously into a higher speed through the broadband wavevector with $\delta k_x c/\omega_0 \gtrsim 0.3$. The staged acceleration of hot electrons leads to the turbulence of electron plasma wave, which is also found in the simulations regarding a lower density plasma [33]. Therefore, large numbers of electrons can be heated to a higher temperature.

The wavevector distributions of the electrostatic wave obtained from p-polarized simulations, shown in Figs. 7(g) and 7(h), exhibit similar features. The white dashed hyperbola and circle denote the expected wavevector distribution corresponding to the maximum growth rate of TPD and the dispersion relation of electrostatic waves, respectively. The significant mitigation of SBS induced by the broadband laser results in more intense absolute SRS and TPD at 0.6% bandwidth. Moreover, the low-coherence laser broadens the density region of absolute SRS and TPD. The variation of the speckle distribution over time of low-coherence laser is slower than the growth rates of SRS and TPD, and thus its effect on SRS and TPD is weaker than that on SBS [50]. As a result, SRS and TPD are significantly enhanced with the 0.6% broadband laser.

In Fig. 8, we present the simulated transverse distributions of SRS and SBS reflectivity for s-polarized lasers with different bandwidths. The overall SBS reflectivity reaches 27.29%, 10.89%, and 1.37% for the narrowband laser, the 0.2% broadband laser, and the 0.6% broadband laser, respectively, indicating that SBS is significantly suppressed with increasing laser bandwidth.

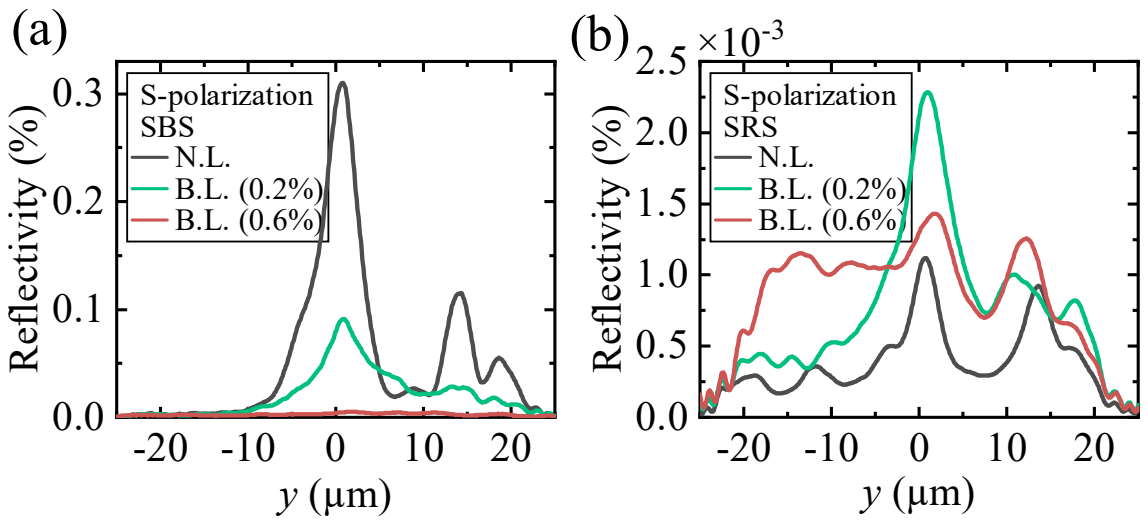

**Figure 8**. Simulated transverse reflectivity distributions for s-polarized lasers with different bandwidths. The distributions of reflectivity corresponding to SBS (a) and SRS (b) were are diagnosed in the left vacuum region.

Due to the absence of a plasma flow profile and the exclusion of the critical density region in the present simulations, the modeling of SBS is unlikely to be quantitatively accurate, as it cannot capture the SBS seed wave originating from unabsorbed light reflected from the critical surface. Accordingly, the discussion of SBS in this work should be regarded as qualitative only. For SRS, the overall reflectivity is 0.19%, 0.38%, and 0.43%, respectively. With the mitigation of SBS, the overall SRS reflectivity continuously increases as the laser bandwidth increases. It can be observed from Fig. 8(b) that for narrowband laser and 0.2% broadband laser, SRS is mainly excited in the intense speckle regions. For the 0.6% broadband laser, the spot profile is more smoothed by CPP, leading to intense SRS across the entire region. Accordingly, the suppression of SBS combined with the intrinsic effect of laser bandwidth results in the increase in SRS reflectivity.

## V. Conclusions

In this manuscript, the characteristic of hot electron generated from plasmas subject to low-coherence lasers was experimentally investigated. The results show a significant enhancement of electron temperature and electron number when using a low-coherence lasers with moderate bandwidth. The energy fraction of hot electrons for the 0.6% bandwidth is estimated to be up to 2.8% based on the measured hot electron and hard X-ray spectra as well as the Monte Carlo simulations, which is approximate 4 times as high as the narrowband cases. The enhancement may be partially attributed to the mitigated SBS and the increased SRS under moderate bandwidths according to the spectral measurement of scatterings. Scattering diagnostics also indicate that the bandwidth intrinsic effect may play a dominant role in the generation of hot electrons. 2-D PIC simulations demonstrate that a broadband laser can develop absolute instabilities in a wide density region, thereby enhancing hot electron generation. Moreover, a broadband laser leads to plasma turbulence with a broad wavevector spectrum, which accelerates electrons continuously to a higher temperature. The findings indicate that moderate-bandwidth lasers can serve as efficient hot-electron sources. In electron shock ignition ICF, the enhancement of hot electrons produced by moderate-bandwidth lasers may strengthen shock waves to reduce ignition energy requirements, if the preheating of cold fuel by the hot electrons can be overcome. Further broadband effect studies at higher laser intensities (~$10^{16}$ W/cm$^2$) are needed to evaluate their impact on shock ignition performance. Moreover, such lasers can generate high-brightness X-ray sources with a 5–8-fold intensity enhancement, thereby improving backlighting and imaging performance for diagnostic applications in ICF and HEDP research.

## Acknowledgments

We acknowledge all the technical support of the engineering staff at Shanghai Institute of Laser Plasma for the laser operation. This work was supported by National Key R&D Program of China (Grants No. 2025YFF0514300), the National Natural Science Foundation of China (Grants No. 11905280 and No. 12005287), the Guangdong Basic and Applied Basic Research Foundation (Grant No. 2023A1515011695), and Instrument Project of the Chinese Academy of Sciences (Grant No. PTYQ2024TD0028). The EPOCH code used in this work was in part funded by the UK EPSRC Grants No. EP/G054950/1, No. EP/G056803/1, No. EP/G055165/1, No. EP/M022463/1, and No. EP/P02212X/1.

## Data availability

The data that support the findings of this study are available from authors upon reasonable request.

## Appendix. Laser conditions and simulation parameters

### A.1. Kunwu laser system.

The Kunwu laser system is described thoroughly in refs. [24,25]. This laser is the world's first kilojoule-class broadband low-coherence Nd:glass driver, which was built to suppress or eliminate the laser-plasma instabilities that arise in attempts to achieve inertial confinement fusion.

The light source employs an Yb-doped fiber pulse amplifier seeded by a super luminescent diode centered at 1050 nm with 50-nm bandwidth. To suppress amplified spontaneous emission, a 25-nm bandpass filter and acousto-optic modulators are integrated into the fiber amplification stage. The spectrum of the source was precisely shaped with a birefringent filter to compensate for the gain narrowing of the following Nd:glass amplifiers. A Pockels cell converts the pulse train to single-shot operation before injection into the preamplifier unit. The pulse is then amplified by Nd:glass rod amplifiers, delivering up to 35 J energy at nanosecond-scale durations with 15-nm bandwidth. The main amplifier adopts a six-stage, five-pass Nd:glass configuration, scaling the broadband pulse to 1 kJ/3 ns while maintaining 13-nm bandwidth.

Due to the advantage of short wavelength in increasing ablation pressure and suppresses laser-plasma instabilities, as well as balancing the output energy and bandwidth, the Kunwu laser was selected to operate at the second harmonic (0.53 μm). A 200 mm × 200 mm aperture type-I DKDP crystal achieves 63% conversion efficiency at its retracing point, with an acceptance bandwidth of ~15 nm. At the terminal, a continuous phase plate is utilized for smoothing focal spot and size adjustment.

The Kunwu laser integrates a switchable seed source, enabling rapid reconfiguration between broadband (15-nm bandwidth) and narrowband (<0.1-nm bandwidth) operational modes, as validated in our experiments.

### A.2. Monte Carlo simulations.

MC simulations have been performed with *EGS*nrc code. The physics libraries selected for Bremsstrahlung cross sections and electron impact ionization were NIST [55] and Penelope [56], respectively. The target used in the simulations had the same parameters as that in the experiment, and the input electron beam was set with the following parameters. In the space domain, the electron beam was set to be circular with a gaussian radial density distribution function of $f(r) \propto \exp(-r^2/2\sigma^2)$, where $\sigma = 100$ μm to match the laser beam diameter in the low-density plasma where hot electrons are produced. In the velocity domain, both the magnitude (corresponding to energy spectrum) and direction are crucial to the simulation. The 3-D Maxwellian fitting distribution of the hot electron spectrum measured by ESM-B was served as the input spectrum. To replicate the wide divergence of hot electron emission [57], the direction of the input electrons was uniformly distributed within an angle of 120° axisymmetric to the target normal. Each simulation involved 2 billion electrons, which was found to be sufficient by testing with more input electrons.

### A.3. Radiation hydrodynamic simulations.

The radiation hydrodynamics code FLASH was employed to characterize the coronal plasma conditions during the experiments. Simulations were conducted in a 2-D axisymmetric geometry using cylindrical coordinates, with the symmetry axis aligned with the optical axis of the incident laser beam, which is perpendicular to the target surface. The incident laser beam was defined and propagated in three dimensions via a 3D-in-2D ray-tracing method. The temporal and spatial intensity profiles were determined based on the measured laser pulse envelope and focal spot characteristics. The energy deposition of the laser in the corona was modeled using an inverse Bremsstrahlung approach, while thermal conduction in the CH plasma was treated with a flux limiter of 0.06. For subsequent analysis, the plasma conditions at mid-pulse time were selected as representative of typical experimental conditions.

### A.4. Particle in cell simulations.

According to the radiation hydrodynamic results, PIC simulations have been performed with 2-D EPOCH code using smoothed laser models. As a widely used kinetic code for investigating the generation of hot electrons, the spatial axis is normalized to the wavelength ($\lambda_0 = 0.527$ μm). Since this study primarily focuses on the effect of laser bandwidth on hot electron generation through non-collisional wave particle interactions, the simulations did not include plasma flow or collisions. The simulation duration of laser-plasma interactions is up to 17.57 ps, when time integral of electron spectra are almost saturated. Considering the typical parametric instabilities for the generation of hot electrons at normal incidence, the plasma density with a finite range $[0.067, 0.31]$ $n_c$ is used in the simulations. Closely approximating actual experimental conditions, the focal spot of low-coherence lasers is a superposition of multi-color beamlets, leading to the speckle distribution changes randomly and rapidly over time. The spatiotemporal variation of speckles can reduce the intensity of parametric instabilities, which has been considered in the presented simulations.

*These authors contributed equally to this work.

†Corresponding author: kangning@siom.ac.cn

‡Corresponding author: lal@siom.ac.cn